# Stress fluctuations in granular matter:

## Normal *vs.* seismic regimes in uniaxial compression tests

### F. Adjémian & P. Evesque

Lab MSSMat, UMR 8579 CNRS, Ecole Centrale Paris
92295 CHATENAY-MALABRY, France, e-mail: evesque@mssmat.ecp.fr

**Abstract:**

This paper is concerned with the trends of stress fluctuations in dry granular materials as functions of the sample size D and of the grain diameter d. Results are obtained in the plateau regime of large axial deformation $e_1$, during axi-symmetric uni-axial compression. Two cases have to be distinguished depending on whether the material generates or does not generate stick-slip.

When no stick-slip is generated (Fig.1), it is found that the relative stress fluctuations $\delta q/q$ of the deviatoric stress $q=\sigma_1-\sigma_3$ scales linearly with the ratio d/D. $\alpha$ ranges from 1 to 3 about, depending on the grain size d; this is probably due to the effect of the piston roughness. This result is compatible with a contact force network with a random force distribution and a short range correlation length $\xi$, i.e. $\xi<3d$; this leads to estimate the representative elementary volume REV to contain few grains (from 1 to 30) as it has been defended already in [1].

We turn now to cases exhibiting stick-slip (Fig. 2). It has been shown in [2] that the distribution of delays between events is rather "exponential" when the sample is small, and Gaussian-like when the sample is larger. The cross-over between the random to quasi periodic regime occurs for a typical sample which contains $10^7$ grains [2]. In this paper we show an analogy between the statistics of stick-slip amplitude $\Delta q/q$ and the seismicity of the Los Angeles area during 1979-1997 period [3]. Similarity of the two trends is quite surprising. The seismicity has been cut off at Magnitude M=3.5.

These results have to be handled with care since similarity may be fortuitous.

**Pacs # :** 5.40 ; 45.70 ; 62.20 ; 83.70.Fn

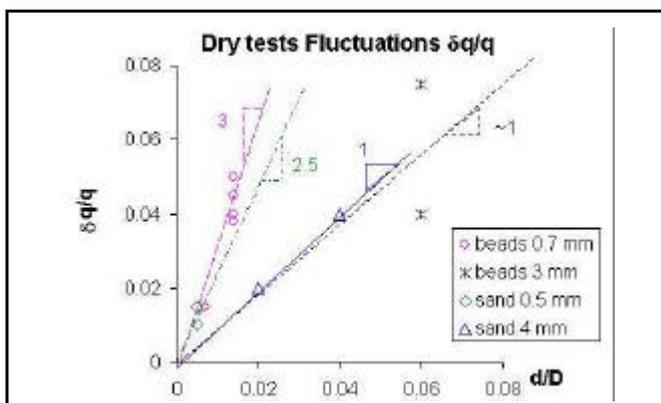
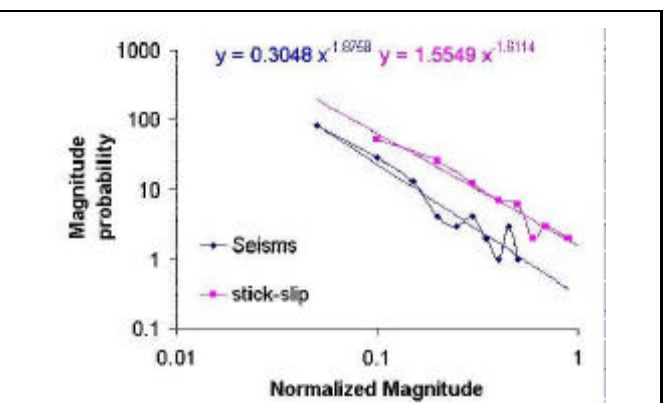

*Fig. 1: Non stick-slip behaviour:* Relative width of the deviatoric stress fluctuations $\delta q/q$ vs. the ratio d/D of the grain size d to the sample diameter D. Trend of $\delta q/q$ is compatible with a linear variation $\delta q/q=\alpha$ (d/D).

*Fig. 2: Stick-slip behaviour* pink data: $\Delta q/q$ distribution in a small sample of glass beads exhibiting stick-slip (from test #29 of ref. [2]) vs. Magnitude distribution (M>3.5) in the LA area during 1979-1997 (blue-black data, from ref. [3]).

**In the non stick-slip case,** the linear trend $\delta q/q = \alpha$ (d/D) exhibited by Fig. 1 depends on the grain diameter d; it is probably due to the roughness of the piston which increases the contact surface when small grains are used. As $\alpha$ is 1-to-3, the estimation of the REV length $\xi$ from $\delta q/q \cong (\xi/D)$ (in 3 dimension) gives $\xi/d$=1-to-3 so that the REV contain from 1-to-30 grains.

**Case with stick-slip:** In paper [2], stick-slips generation has been studied in axial compression test of some samples of glass spheres. It has been found that (i) stick-slips occur in samples which exhibit a weakening of the stress-strain characteristics when the axial strain rate $\partial\varepsilon_1/\partial t$ is increased, that (ii) statistics of delays between successive stick-slips change of behaviours when increasing the sample size (and the strain rate); this distribution passes from some "exponential" distribution to a Gaussian one when sample diameter passes from D=5cm to D=10cm. The cross-over from "exponential" to Gaussian characterises a strong modification of stick-slip production from erratic and random to quasi-periodic. This cross-over defines the REV size to obtain macroscopic regime, which turns out to contain $10^7$ grains=REV!!! In Fig. 2 the distribution of stress falls $\Delta q/q$ during stick-slip has been plotted when the sample is smaller than the REV, *i.e.* experiment #29 of [2]. In the terminology of critical phenomena it corresponds then to a sample smaller than the correlation length, and hence to finite size effect. So the parallel with finite size effect in avalanching and with 1/f noise problem [4,5] is obvious. Experiment trends of Held et al. was already interpreted by one of us [6] using classical critical-subcritical bifurcation theory and classic soil mechanics concepts. Liu & Nagel interpretation [7] came later.

Power law fits lead to confidence factors $R^2$=0.95 (& $R^2$=0.92) for stick-slip data (& LA seism data); while an exponential fit leads to $R^2$=0.88 ($R^2$=0.81) respectively. This makes the power law a better fit than the exponential. However these results have to be taken with much care and have to be confirmed prior to assert it is the first step for a lab modelling of seismic activity [8].

This article is now open to discussion; remarks have to be sent as full signed paper to *poudres & grains.*

*Acknoledgment:* CNES is thanked for partial funding.

## References

[1] P. Evesque, "Macroscopic Continuous Approach versus Discrete Approach, Fluctuations, criticality and SOC", *poudres & grains* **12** (8), 122-150 ( Novembre 2001)

[2] F. Adjémian & P. Evesque, "Experimental stick-slip behaviour in triaxial test on granular matter", *Poudres & Grains* **12** (7), 115-121, (2001), ISSN 1257-3957, http://prunier.mss.ecp.fr /poudres&grains/ poudres-index.htm

[3] from: http://quake.geo.berkeley.edu/ncedc/catalog-search.html

[4] P. Bak, C. Tang & K. Wiesenfeld, *Phys. Rev. Lett.* **59**, 381, (1987); *Phys. Rev.* **A38**, 364, (1988); C. Tang, & P. Bak, *Phys. Rev. Lett.* **60**, 2347, (1989)

[5] G.A. Held, D.H. Solina II, D.T. Keane, W.J. Haag, P.M. Horn & G. Grinstein, *Phys. Rev. Lett.* **65**, 1120 (1990)

[6] P. Evesque, "Analysis of processes governing sandpile avalanches using soil mechanics results and concepts.", *Phys. Rev.* **A 43** , 2720, (1991); *J. Physique* (Paris) **51**, 2515 (1990).

[7] C.H. Liu & S.R. Nagel, *Phys. Rev. Lett.* **68**, 2301-2304, (1992) (anteriority of ref. [6] compared to [7] has been admitted at least by one of the author of ref. [7] in a referee report of an rejected comment submitted to Phys. Rev. Lett.)

[8] L. Gil and D. Sornette, *Phys. Rev.Lett.* **76**, 3991-3994 (1996)